\documentstyle[prl,aps,multicol,epsfig]{revtex}

\title{Period halving of Persistent Currents in Mesoscopic M\"{o}bius ladders}
\author{ DENG Wen-Ji$^{1}$,  XU Ji-Huan$^{1}$,  LIU Ping$^{2}$}
\address{$^{1}$Department of Physics, South China University of Technology, Guangzhou 510641, China;\\
$^{2}$Research Institute of Materials Science, South China University of Technology, Guangzhou 510641, China.}

\begin{document}

\maketitle

\begin{abstract}
We investigate the period halving of persistent currents(PCs) of
non-interacting electrons in isolated mesoscopic M\"{o}bius
ladders  without disorder, pierced by Aharonov-Bhom flux. The
mechanisms of the period halving effect depend on the parity of
the number of electrons as well as on the interchain hopping.
Although the data of PCs in mesoscopic systems are
sample-specific, some simple rules are found in the canonical
ensemble average, such as all the odd harmonics of the PCs
disappear, and the signals of even harmonics are non-negative.
\end{abstract}

\begin{multicols}{2}
The persistent currents(PCs) in mesoscopic rings has been explored
for many years\cite{Imry}. Early in 1983, the effect of elastic
scattering was first understood by B\"{u}ttiker {\it et al.},  and
thus the possibility of observing nano-amper PCs in nano-scale
normal metal rings with disorder was revealed\cite{Buttiker1}.
Since then the problem of PCs has been studied in many distinct
aspects\cite{Buttiker2,Cheung,Paris,exp1,exp2,exp3,exp4,exp5,Abraham,kato,Zhu,Zhou,Wang1,Wang2,Chen,Mobius,Wang3,Deng1,Deng2}.
Among the interesting characters of PCs, the period halving
phenomenon is particular simple but significant. In the early
works about the ring systems with Aharonov-Bohm(AB)
flux\cite{Bayers,Bloch}, it has been shown generally that all
relevant physical properties of this ``ring'' are periodic in
$\Phi$ with period of one flux quantum $\Phi_{0}\equiv h/e$, here
$h$ and $e$ denotes the Planck constant and the charge of
electron, respectively. However the studies on PCs in canonical
ensemble find that the fundamental harmonic of this current (with
flux period $\Phi_0$) is strongly suppressed, whereas the next
harmonic (with flux period $\Phi_0/2$)
survives\cite{Cheung,Paris}. Recently the phenomena of PCs  are
studied in some special systems, such as in the M\"{o}bius
ladder\cite{Mobius} as well as in double-ring
structures\cite{Wang3}. In the present work, we  study the systems
of non-interacting electrons in isolated mesoscopic M\"{o}bius
ladders. After re-building the general model of M\"{o}bius ladder
systems, we prove that the well-known formula for PCs in
one-dimensional(1D) rings\cite{Buttiker1} still holds for these
complex structures. Then a series of simulation results of PCs
versus the AB flux and relevant Fourier harmonics are presented,
which manifest the complicated ways in which the interchain
hopping elements and the parity of the number of total electrons
influence the periodic behavior of PCs.

The model of M\"{o}bius ladder originates from a mathematical
concept of non-orientable surface, i.e., the M\"{o}bius band (or
strip), which can be formed by joining the ends of a rectangle
with one twist of 180 degrees\cite{Mobius}. The schematic drawing
of a M\"{o}bius ladder with $N$ rungs is shown in Fig.1. The two
chains on the opposite sides of the rectangle are connected as the
unique edge of the M\"{o}bius ladder. If all the rungs are Broken
off, the ladder will become a double-ring strcture\cite{Wang3}.
For the sake of simplicity, we take the  energy integral between
two nearest-neighbor sites along the edge as the energy unit, then
the Hamiltonian for electrons with spin $\sigma$ (up or down) in
such a lattice can be written as
\begin{eqnarray}\label{eqH2}
\hat{H}_{\sigma}&=&-\sum_{n=1}^{2N}\left(e^{i\theta}\hat{a}_{n+1\sigma}^{\dagger}\hat{a}_{n\sigma}+e^{-i\theta}\hat{a}_{n\sigma}^{\dagger}\hat{a}_{n+1\sigma}\right ) \nonumber\\
&&-t_{\bot}\sum_{n=1}^{N}\left(\hat{a}_{n+N\sigma}^{\dagger}\hat{a}_{n\sigma}+\hat{a}_{n\sigma}^{\dagger}\hat{a}_{n+N\sigma}\right )
\end{eqnarray}
where $\hat{a}_{n\sigma}$ ($\hat{a}_{n\sigma}^{\dagger}$) is an annihilation (creation) operator for an electron with spin $\sigma$ at site $n$, the phase $\theta\equiv2\pi\Phi/N\Phi_{0}$ arises from the path integral of vector potential $\int\vec{A}\cdot d\vec{r}$ between corresponding two nearest-neighbor sites along the edge of M\"{o}bius ladder, and $\Phi$ is the AB flux through the M\"{o}bius structure.
In addition, the on-site energy for each site is set to be zero, but the matrix element $t_{\bot}$ for interchain hopping, i.e., hopping perpendicular to the edge is an important tunable parameter of the systems.
The two kinds of electrons with spin-up and spin-down  will contributes to the total Hamiltonian in a simple way that $\hat{H}=\hat{H}_{\uparrow}+\hat{H}_{\downarrow}$.

Such a Hamiltonian is readily diagonalized as
\begin{equation}
\hat{H}=\sum_{k,\sigma}\varepsilon_{k}\hat{b}_{k\sigma}^{\dagger}\hat{b}_{k\sigma}
\end{equation}
with eigen-energy values
\begin{equation}\label{eq6}
\varepsilon_{k}=-2\cos\frac{\pi}{N}\left (k+\frac{2\Phi}{\Phi_0} \right )-t_{\bot}(-1)^k
\end{equation}
by introducing an unitary transformation
\begin{equation}\label{eq3}
\hat{b}_{k}\equiv \frac{1}{\sqrt{2N}}\sum_{n=1}^{2N}e^{ikn}\hat{a}_{n},\ \ \hat{a}_{n}= \frac{1}{\sqrt{2N}}\sum_{k}e^{-ikn}\hat{b}_{k}
\end{equation}
here $k=0,1,2,\cdots,2N-1$ so that the wave vectors $k\pi/N$ is limited in the first Brillouin zone and the periodic condition $\hat{a}_{n+2N}=\hat{a}_{n}$ is satisfied.
If $t_{\bot}>2$, then Eq.(\ref{eq6}) depicts a two-band spectrum with band gap $\Delta E=2(t_{\bot}-2)$.

At first glance, any single eigen-energy given by Eq.(\ref{eq6}) is a periodic function of $\Phi$ with period $N\Phi_0$, but in fact the PCs vary with the AB flux with period $\Phi_0$ because  thermodynamic properties are determined by the whole eigen-energy spectrum.
The periodicity of  $\{\epsilon_k\}$ is easy to see by using an equality  $\varepsilon_k(\Phi+\Phi_0)=\varepsilon_{k+2}E(\Phi)$.
This is very the reason why all the calculations of PCs can be limited in the flux range $\Phi\in[0,\Phi_0)$.

However, a M\"{o}bius ladder with  $t_{\bot}=0$ becomes
a double-ring structure\cite{Wang3}, and the dispersion relation Eq.(\ref{eq6}) will be reduced as
\begin{equation}\label{eq5}
\varepsilon_{k}=-2\cos\frac{\pi}{N}\left ( k+\frac{\Phi^{\star}}{\Phi_0} \right )
\end{equation}
here $\Phi^{\star}\equiv 2\Phi$ is defined as the flux-linkage
through the double-ring structure. The flux-linkage is two times
of the AB flux because the electrons have to travel along the full
edge of the M\"{o}bius ladder, thus encircling the flux twice.
Therefore the PCs in a double-ring system  must be a periodic
function of AB flux with period half a flux quantum $\Phi_0/2$
because of
$\varepsilon_{k}(\Phi^{\star}+\Phi_0)=\varepsilon_{k+1}(\Phi^{\star})$.
In Ref.\cite{Wang3}, this simple result is referred as ``Entire
period halving '' effect.

There is no obvious reason to ensure that the well-known formula
of PCs in one-dimensional ring systems is still held for some
subtle cases such as the M\"{o}bius ladders. Therefore we try to
prove it based upon some basic concept of quantum mechanics. It is
obvious that the current at $nth$ rungs involve two parts: the
current from $n$ to $n+1$ in one chain and the current from $n+N$
to $n+N+1$ in another chain. So we have
\begin{eqnarray}\label{eq8}
\hat{\mathcal{J}}&=&\hat{\jmath}_{n}+\hat{\jmath}_{n+N} \nonumber\\
&=&\frac{1}{N}\sum_{n=1}^{N}(\hat{\jmath}_{n}+\hat{\jmath}_{n+N}) \nonumber\\
&=&\frac{1}{N}\sum_{n=1}^{2N}\hat{\jmath}_{n}
\end{eqnarray}
in the second step of above derivation, the conservation condition of current has been used. Applying the concept of probability current in Hilbert space\cite{Deng3} \begin{eqnarray}\label{eq9}
\hat{\jmath}_{n}&=&\hat{\jmath}_{n\rightarrow n+1}-
\hat{\jmath}_{n+1\rightarrow n} \nonumber \\
&=&\frac{1}{i\hbar}\sum_{\sigma}\left(e^{i\theta}\hat{a}_{n+1\sigma}^{\dagger}\hat{a}_{n\sigma}-e^{-i\theta}\hat{a}_{n\sigma}^{\dagger}\hat{a}_{n+1\sigma}\right )
\end{eqnarray}
to above equation (\ref{eq8}), then we prove that the charge current operator for the Hamiltonian system defined in Eq.(\ref{eqH2}) can still be re-obtained as $\hat{I}(\Phi)=-\partial \hat{H}(\Phi)/\partial \Phi $. Simple calcualtion yields the formula of PCs in a M\"{o}bius ladder
\begin{equation}\label{eq10}
I(\Phi)=-e\left < \hat{\mathcal{J}} \right>
=-I_0\sum_{k,\sigma}\overline{n}_{k\sigma}\sin\frac{\pi}{N}\left (k+\frac{2\Phi}{\Phi_0}\right)
\end{equation}
where $I_0\equiv 4\pi/N\Phi_{0}$, and $\overline{n}_{k\sigma}(\Phi)\equiv\left < \hat{b}_{k\sigma}^{\dagger}\hat{b}_{k\sigma} \right > $
stands for the thermodynamic averaged number of electrons in quantum state $(k,\sigma)$.

The formula Eq.(\ref{eq10}) provide us a base to study the PCs in M\"{o}bius ladders.
For the sake of clarity, we will in the present paper focus on the systems with fixed number of electrons at absolute zero temperature.
In this case, all the $N_e$ electrons will fill into single-electron quantum states $(k,\sigma)$ according to their eigen-energy values from ground state to higher energy states one by one.
It means that $\overline{n}_{k\sigma}=1$ for occupied states and otherwise $\overline{n}_{k\sigma}=0$.

The conventional way for analyzing the periodic behavior of PCs is
to calculate the Fourier harmonics of the data of currents at $K$
discrete values of AB flux $\Phi_\ell=\ell\Phi_0/K$,
$\ell=0,1,\cdots,K-1$. Choosing the basis functions as
$\varphi_{m,\ell}=K^{-1/2}\exp (i\omega_m \ell/K)$ with
$\omega_m\equiv 2m\pi$, then we have
\begin{mathletters}
\begin{equation}\label{I}
I_\ell=\frac{1}{\sqrt{K}}\sum_{m=0}^{K-1}F_{m}\exp(i\ell\omega_m/K),
\end{equation}
\begin{equation}
F_m=\frac{1}{\sqrt{K}}\sum_{\ell=0}^{K-1}I_{\ell}\exp(-i\ell\omega_m/K)
\end{equation}
\end{mathletters}
for $m, \ell=0,1,\cdots,K-1$, and the spectra theorem is written as
$\sum_{m=0}^{K-1}|F_{m}|^{2}=\sum_{\ell=0}^{K-1}|I_{\ell}|^{2}$.
In fact, the real part of $F_m$ keeps to be zero and $F_{K-m}=F_{m}^{*}$ because $I(\Phi_0-\Phi)=-I(\Phi)$ for $0\leq\Phi<\Phi_0$. Thus  Eq.(\ref{I}) can be re-written as
\begin{equation}
I_\ell=\frac{1}{\sqrt{K}}\sum_{m=1}^{K-1}S_{m}\sin(\ell\omega_m/K),
\end{equation}
here $S_m=-Im(F_m)$ and $Im(\cdots)$ denotes the imaginary part of a complex number.
In practical experiments\cite{exp1,exp2,exp3,exp4,exp5}, $S_m$ will be detected as the amplitude of signals of harmonics $\omega_m=2m\pi$ with period $\Phi_0/m$. In what follows, we will present only first twenty harmonics\cite{Sm}.

The plots in Fig.2 are some typical results selected from our numerical simulations on PCs in double-ring systems.
We find that the PCs in any single double-ring system with odd number of electrons is a period function of $\Phi$ with period $\Phi_0/4$, halving of the period $\Phi_0/2$ as discussed in the paragraph below eq.(\ref{eq5}) and shown in Fig.2(a), and that
the second harmonic with period $\Phi_0/2$ of even number electrons in single double-ring system doesnot disappear as shown in Fig2.(b); but however we find  also that the second harmonic will be strongly suppressed in summation of PCs in two systems with adjacent even numbers of electrons, such as shown Fig.2(c).
Note that a double-ring system is equivalent to a 1D ring pierced by a AB flux $\Phi^{\star}\equiv 2\Phi$, we suppose that the parity effect of electron number exists in the PCs of mesoscopic rings.
We hope that this observation will shed light on the mechanism of period behaving of PCs.

Some other interesting properties of PCs in canonical ensemble
average are shown in Fig.3. It is well-known that the PCs in
mesoscopic systems are sample-specific, but simple rules always
appear in some kinds of averages. We calculate the PCs and
relevant Fourier harmonics varying with the AB flux for different
M\"{o}bius ladders with different fixed numbers of electrons. The
data of single system show that the PCs can be paramagnetic or
diamagnetic, the first harmonic can exist or vanish. Following the
works of Bouchiat and Montambauxa\cite{Paris}, we take the
arithmetic mean values od PCs on the number of electrons. The
averaged PCs and corresponding harmonics manifest some surprising
characters. All the odd harmonics disappear, and the even
harmonics keeps to be non-negative. In addition, very small
probability amplitude of inter-chain hopping, as small as
$t_{\bot}=0.01$ shown in Fig.3(b), can result in a considerable
harmonic of period $\Phi_0/2$.

In conclusion we study the problem of PCs in M\"{o}bius ladder
systems  of non-interacting electrons with a tuneable parameter
$t_{\bot}$. It is proven exactly that the well-known formula for
PCs in 1D rings still holds for the complicated new structures. A
series of simulation results manifest that the perpendicular
hopping elements and the parity of the number of total electrons
influence the periodic behaviors of PCs dramatically. As proposed
by Mila et al\cite{Mobius}, a M\"{o}bius structure with
controllable inter-chain hopping could be fabricated in GaAs
hetero-structure using a wide quantum wire. Our present study
provides a base for further investigations on the disorder effect,
temperature effect, and the effect of electron interaction. Then
we can expect that these relevant theoretical results are testable
in experiment.

This work is supported by National Key Program for Basic Research, 2001-03500, and partly by the Guangdong Provincial Natural Science Foundation of China.

\end{multicols}

\begin{figure}
\caption{Schematic drawing of a M\"{o}bius ladder with $N$ rungs.
The AB flux $\Phi$ pierces through the center of the
structure.}\label{Fig.1}
\end{figure}
\begin{figure}
\caption{Typical curves of PCs (left column) and first twenty
harmonics of Fourier transformation (right column) for $N=100$
double-ring systems. (a) odd number of electrons $N_e=281$; (b)
even number of electrons $N_e=280$; (c) summation of the PCs and
corresponding harmonics of two systems with adjacent even nembers
of electrons, i.e., $N_e=280$ and $N_e=282$.}\label{Fig.2}
\end{figure}
\begin{figure}
\caption{Typical curves of canonical ensemble averaged PCs (left
column) and first twenty harmonics(right column). (a)
$t_{\bot}=0$; (b) $t_{bot}=0.01$; (c)$t_{\bot}=1.0$.}\label{Fig.3}
\end{figure}
\begin{figure}[btp]
\begin{center}
\leavevmode \epsfbox{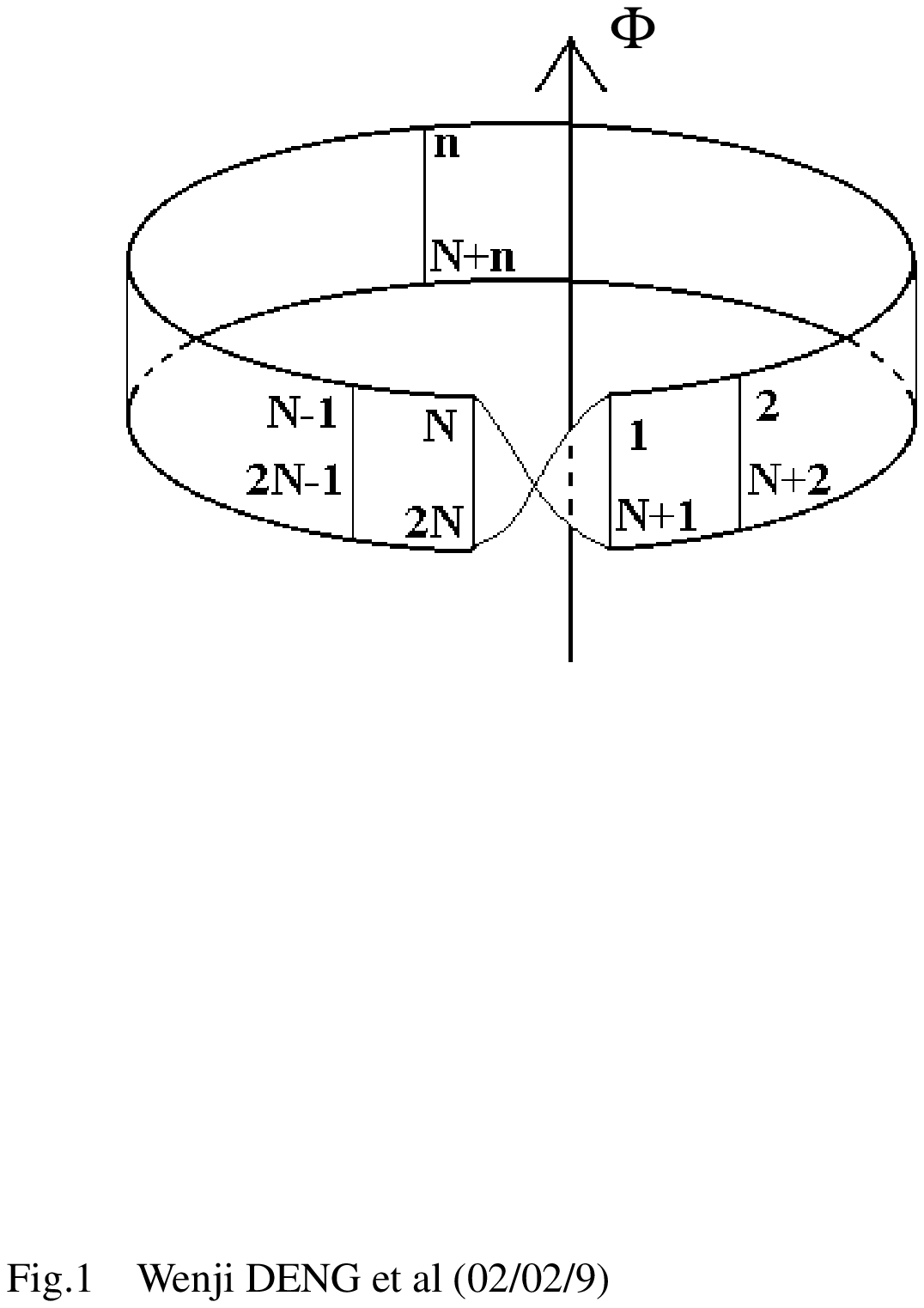}
\end{center}
\end{figure}
\begin{figure}[btp]
\begin{center}
\leavevmode \epsfbox{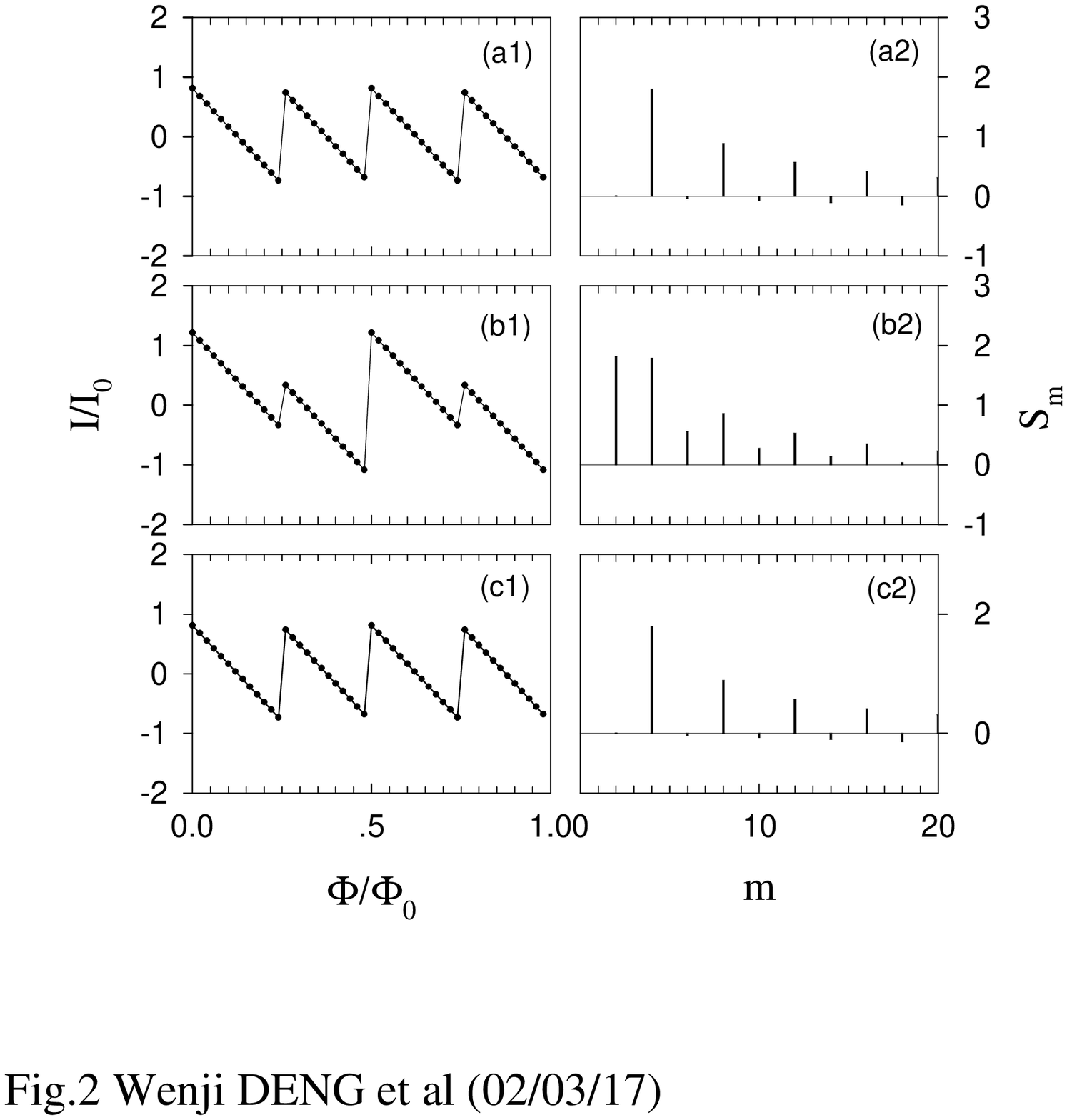}
\end{center}
\end{figure}
\begin{figure}[btp]
\begin{center}
\leavevmode \epsfbox{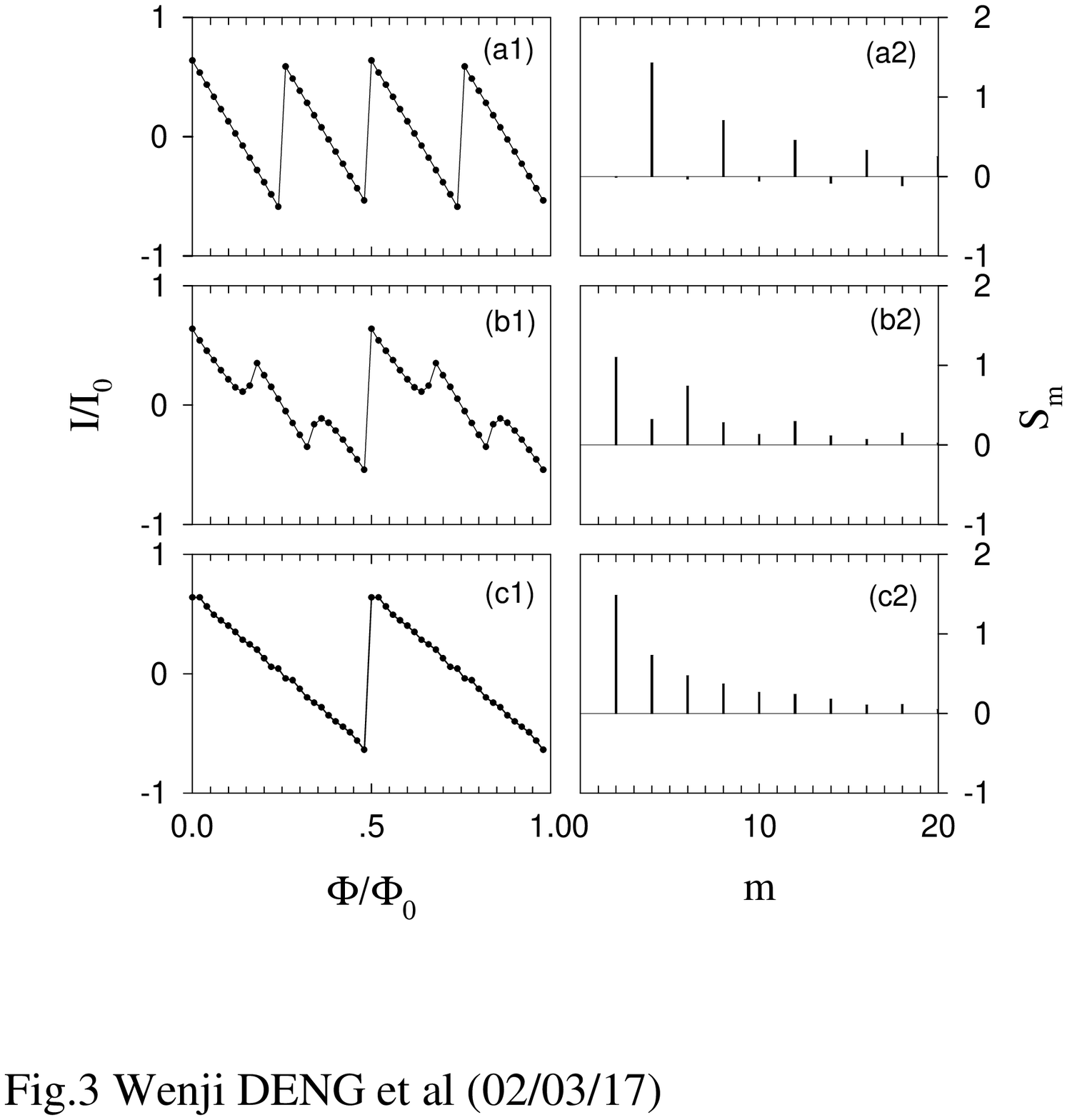}
\end{center}
\end{figure}

\begin{references}
\bibitem{Sm}Strictly speaking, $S_m$ is not the amplitude of $m$th harmonic; but for the low order harmonics, it is a very good approximation.
\end{references}
\end{document}